\documentclass[aip,jcp,twocolumn,showpacs,superscriptaddress,amsmath,amssymb,floatfix,10pt]{revtex4-1}  
\usepackage{graphicx}  
\usepackage{dcolumn}   
\usepackage{bm}        
\usepackage{amssymb}   
\usepackage{framed}
\usepackage{amsmath}
\usepackage{multirow}
\usepackage{hhline}
\usepackage{color}
\definecolor{bluegreen}{rgb}{0,0.2,0.8}
\usepackage{subfigure,amsmath,verbatim,moreverb}
\usepackage{tabularx}
\usepackage{adjustbox}
\usepackage{lipsum}

\begin{document}

\widetext
\title{Employing semilocal exchange hole with an application to meta-GGA level screened range separated hybrid functional: A conventional wisdom method}

\author{Subrata Jana}
\email{subrata.jana@niser.ac.in}
\affiliation{School of Physical Sciences, National Institute of Science Education and Research, HBNI, Bhubaneswar 752050, India}
\author{Prasanjit Samal}
\email{psamal@niser.ac.in}
\affiliation{School of Physical Sciences, National Institute of Science Education and Research, HBNI, Bhubaneswar 752050, India}

\date{\today}

\begin{abstract}

Range separated hybrid density functionals are very successful in describing a wide range of molecular and solid state properties accurately. 
Range separated hybrid functionals are designed from spherically averaged or system averaged reversed engineered exchange hole. In the 
present attempt, we employ screened range separated hybrid functional scheme to the meta-GGA rung by using Tao-Mo semilocal exchange 
hole (or functional). The hybrid functional proposed here utilizes the spherically averaged density matrix expansion based exchange 
hole in range separation scheme. For slowly varying density correction, we employ range separation scheme only through the local 
density approximation (LDA) based exchange hole coupled with the slowly varying Tao-Mo enhancement factor through the conventional 
wisdom technique. Comprehensive performance and testing of the present functional shows, it accurately describes several molecular 
properties. The most appealing feature of this present screened hybrid functional is that it will be practically very useful in 
describing solid state properties in meta-GGA level.   

\end{abstract}

\maketitle

 Kohn-Sham variant of density functional theory (DFT)~\cite{ks65} is one of the most accurate and widely used many body frameworks 
 for electronic structure calculation. The theoretical framework of KS DFT is exact, only if the exact form of exchange-correlation 
 (XC) is known. The accuracy of DFT depends upon the accuracy of XC functional, which contains all the many electron effects. 
 Therefore, development of accurate XC functional is an intriguing research topic with different new prospects. A large number 
 of approximations based on different global models have been developed in recent decades~\cite{PW86,B88,LYP88,BR89,PW91,
B3PW91,B3LYP,PBE96,VSXC98,HCTH,PBE0,HSE03,AE05,ZWu06,MO6L,TPSS03,revTPSS,PBEsol,SCAN15,Kaup14,Tao-Mo16,con1,con2,con3,con4}.  
All these approximations are classified through Jacob’s Ladder approximations, which add an extra ingredient in each of its rung 
starting from the local density approximation (LDA). The generalized gradient approximations (GGAs)~\cite{PBE96,PW91,ZWu06} and 
meta-generalized gradient approximation (meta-GGA)~\cite{TPSS03,MO6L,SCAN15,Tao-Mo16} are the next two higher rungs of Jacob's 
ladder. LDA, GGA and meta-GGA are  enormously used due to their low computational cost and accuracy in chemistry~\cite{CJCramer09,
Quest12,Yangreview,Becke14,VNS03,PHao13,LGoerigk10,LGoerigk11} and condensed matter physics~\cite{YMo16,Csonka09,PBlaha09,
FTran16,CAdamo10,VNS04}. All these approximations are known as semilocal formalism because they are constructed from semilocal 
quantities i.e., density ($\rho$), gradient of density ($\nabla\rho$) and Kohn-Sham kinetic energy density ($\tau$). In spite 
of its success of describing several thermochemistry test~\cite{VNS03,PHao13,LGoerigk10,LGoerigk11}, equilibrium lattice constants
~\cite{Csonka09,PBlaha09,FTran16}, bulk modulus~\cite{YMo16}, bond lengths~\cite{CAdamo10,YMo16}, cohesive energy~\cite{VNS04} 
and solid state surface properties ~\cite{YMo16}, semilocal approximations often fail to predict phenomena like solid state 
band gap, thermochemical reaction barrier heights~\cite{vsp07}, excitation energies, Rydberg excitation and dissociation 
curves~\cite{mcy06,rpcvs06}, because of the inherent absence of "non-locality" and "many electron self interaction 
(MESI)"~\cite{vsp07,mcy06,rpcvs06,pz81,prp14,hkkk12,phl84,rpcvs07,pplb82,rmjb17}. Non-locality and MESI information are 
induced inside the semilocal formalism by suitably mixing Hartree-Fock (HF) exact exchange globally or in range 
separated scheme~\cite{B3PW91,B3LYP,PBE0,HSE03,rmjb17,hse06,VNS03,JJaramillo03,camb3lyp,mcy06,lcwpbe,hs10,jks08,whc,
tpssrs}. The former is known as hybrid~\cite{B3LYP,PBE0,VNS03}, whereas the later is known as range separated 
hybrid~\cite{hse06,camb3lyp,mcy06,lcwpbe,rmjb17,tpssrs,savin1,savin2,savin3}. Construction of RS functionals are based on the semilocal 
exchange hole. The exchange hole can be construed either in Taylor series expansion~\cite{BR89} or density matrix 
expansion (DME) technique~\cite{Tao-Mo16} or reversed engineering way~\cite{hse06,lcwpbe,tpssrs,tpsshole,Lucian13}. 
Recently, Tao-Mo~\cite{Tao-Mo16} developed a full semilocal exchange hole based on DME technique having all desired 
properties together. 

Inspired by TM functional~\cite{Tao-Mo16} and its underlying exchange hole, in this present paper we have constructed a 
screened RS hybrid functional using HF in its short range part. The present proposition is the same direction as the 
RS hybrid functional was proposed by Heyd-Scuseria-Ernzerhof (HSE) ~\cite{hse06}, but in the meta-GGA rung. The DME 
based exchange hole is used for constructing the full short or long range part of semilocal exchange functional. For 
slowly varying density correction we impose the range separated scheme only through the LDA exchange hole. The motivation 
of inclusion only LDA exchange hole in the present construction is rooted in the construction of coulomb attenuated B88 
family long range corrected RS functionals~\cite{camb3lyp}. More details and physical inside behind our functional 
construction is discussed in the construction section. To check the accuracy and performance of the present proposition we applied 
our scheme to the well-known test set. Performance of the present functional is compared with three very popular 
hybrids B3LYP, PBE0, TPSSh and one RS hybrid HSE06. All these functionals are designed from their popular parent 
semilocal functionals B88, PBE and TPSS by suitably mixing exact HF exchange. Special attention has been paid to 
the performance of present proposition with that of HSE06 functional for thermochemistry and fractional occupation 
number because both the functionals are designed using HF in its short range. 

The present paper is organized as follows. In the following, we will discuss the role of exchange hole in designing 
the RS hybrids. Next, we will propose our formalism in designing a new meta-GGA level range separated functional. The 
differences and similarities with other RS functionals are also discussed. Lastly, we compare present proposition with 
that of popular hybrids and RS hybrid functional for different thermochemistry test cases.

The exchange hole $\rho_{x}$ is the principle constituent of exchange energy functional. It is used for constructing the 
semilocal exchange energy functional or range separation hybrid functional. The exchange energy functional constructed from 
exchange hole is given by, 
\begin{eqnarray}
E_{x}[n] = \frac{1}{2} \int d^3r~ \rho(\mathbf{r})\int d^3u \frac{\rho_{x}(\mathbf{r},\mathbf{r}+\mathbf{u})}{u}~.
\end{eqnarray}

On the other hand, the range separated density functional theory is developed by separating the Coulomb interaction i.e., 
$v_{ee}({\mathbf{r}}, {\mathbf{r}'})=\frac{1}{|\mathbf{r}-\mathbf{r}'|}$ into short and long range part as,
 \begin{equation}
 \frac{1}{|\mathbf{r}-\mathbf{r}'|}=\underbrace{\frac{erf(\mu |\mathbf{r}-\mathbf{r}'|)}{|\mathbf{r}-\mathbf{r}'|}}_{LR}
 +\underbrace{\frac{1-erf(\mu |\mathbf{r}-\mathbf{r}'|)}{|\mathbf{r}-\mathbf{r}'|}}_{SR}
 \end{equation}
By replacing $\mathbf{r}'=\mathbf{r}+\mathbf{u}$, and using the exchange hole, the short range and long range part of exchange 
functional becomes,
\begin{equation}
 E_x^{SR} = -\frac{1}{2}\int~d^3r\rho(\mathbf{r})\int\frac{1-erf(\mu u)}{u}
 \rho_x(\mathbf{r},\mathbf{r}+\mathbf{u})~d^3u,
\end{equation}
and 
\begin{equation}
 E_x^{LR} = -\frac{1}{2}\int~d^3r\rho(\mathbf{r})\int\frac{erf(\mu u)}{u}
 \rho_x(\mathbf{r},\mathbf{r}+\mathbf{u})~d^3u,
\end{equation}
In the construction of exchange hole, one needs to formulate only the spin-unpolarized exchange hole. Using the spin density 
scaling relation the spin-unpolarized exchange hole can be transformed into spin-polarized form as,
\begin{eqnarray}\label{spinscaling}
\rho_{x}[\rho_\uparrow,\rho_\downarrow] = 
\frac{\rho_\uparrow}{\rho}\rho_{x}[2\rho_\uparrow]+
\frac{\rho_\downarrow}{\rho}\rho_{x}[2\rho_\downarrow].
\end{eqnarray}
It is noteworthy to mentioned that exchange energy depends on the spherical average the exchange hole over separation vector 
$\mathbf{u}$. Hence, Eq.(1) can be rewritten as,
\begin{eqnarray}
E_{x}[n] = \frac{1}{2} \int d^3r~ \rho(\mathbf{r})\int d^3u \frac{\langle\rho_{x}(\mathbf{r},\mathbf{r}+\mathbf{u})\rangle}{u},
\end{eqnarray}
In range separated DFT also the semilocal LR or SR part is constructed from spherical averaged semilocal exchange hole. Not only 
that, the semilocal SR and LR part can also be constructed from reversed engineered system averaged exchange hole. The screened 
hybrid functional HSE~\cite{hse06}, long range corrected LC-$\omega$PBE~\cite{lcwpbe} are designed from reversed engineered exchange hole. 
Also, very recently TPSS exchange energy functional is reversed engineered to construct meta-GGA level screened hybrid functional~\cite{tpssrs}.     


The general scheme of hybrid functional~\cite{B3LYP} is,
\begin{equation}
E^{hybrid}_{xc}= aE_x^{HF} + (1 - a)E^{SL}_x + E^{SL}_c,    
\end{equation}
where $E_x^{HF}$ is the Hartree-Fock exact exchange, $E^{SL}_x$ is the semilocal exchange functional and $E^{SL}_c$ is the semilocal 
correlation functional. Here, $a$ controls the amount of HF to be mixed with the semilocal exchange functional. However, there is no 
mixing parameter associated with correlation because it is essential for both the HF and semilocal exchange.

The RS scheme proposed by Heyd, Scuseria, and Ernzerhof (HSE) has the following general form,
\begin{equation}
    E_{xc} = aE_x^{HF-SR} + (1-a)E^{SL-SR}_x + E^{SL-LR}_x + E^{SL}_c~.  
\end{equation}
Alternatively this can be written as,
\begin{equation}
    E_{xc} = aE_x^{HF-SR} -aE^{SL-SR}_x + E^{SL}_x + E^{SL}_c  
\end{equation}
This is particularly useful in implementation point of view as one has to construct only the short range part of semilocal exchange 
functional. The last two part added to the semilocal XC functional, which in the present case is the TM functional~\cite{Tao-Mo16}. Thus 
knowing the exchange energy functional form, the main aim is only to construct the semilocal short range part.  

Here, we propose following short range semilocal part of the full range separated hybrid functional scheme,

\begin{widetext}
\begin{equation}
\begin{split}
 E_{x}^{SL-SR}=-\int~n(\mathbf{r})\epsilon_x^{unif}\Big[wF_x^{DME-SR}+
 \Big\{1-\frac{8}{3}\tilde{A}\Big(\sqrt{\pi}~erf(\frac{1}{2\tilde{A}})+(2\tilde{A}-4\tilde{A}^3)e^{-\frac{1}{4\tilde{A}^2}}-3\tilde{A}
 +4\tilde{A}^3\Big)\Big\}(1-w)F_x^{TM-sc}\Big]~d^3r,
 \label{eq6}
 \end{split}
 \end{equation}
 where
 \begin{equation}
 \begin{split}
F_x^{DME-SR} = \frac{1}{f^2}\Big\{1-\frac{8}{3}A\Big(\sqrt{\pi}~erf(\frac{1}{2A})+(2A-4A^3)e^{-\frac{1}{4A^2}}-3A+4A^3\Big)\Big\}\\
+\frac{7\mathcal{L}}{9f^4}\Big\{1+24A^2\Big((20A^2-64A^4)e^{-\frac{1}{4A^2}}-3-36A^2\\
+64a^4+10\sqrt{\pi}~erf(\frac{1}{2A})\Big)\Big\}
  +\frac{245\mathcal{M}}{54f^4}\Big\{1+\frac{8}{7}A\Big((-8A+256A^3-576A^5
 +3849A^7-122880A^9)e^{-\frac{1}{4A^2}}\\+24A^3(-35+224A^2-1440A^4+5120A^6)
 +2\sqrt{\pi}(-2+60A^2)erf(\frac{1}{2A})
 \Big)\Big\}
 \end{split}
 \end{equation}
 is the short range enhancement factor based on density matrix expansion based semilocal exchange hole~\cite{bpatra} and
 \begin{equation}
 F_x^{TM-sc}=\Big[1+10\Big\{\Big(\frac{10}{81}+\frac{50p}{729}\Big)p+
 \frac{146}{2025}\tilde{q}^2-\Big(\frac{73\tilde{q}}{405}\Big)\Big[\frac{3\tau^w}
 {5\tau}\Big](1-\frac{\tau^w}{\tau}\Big)\Big\}\Big]^{\frac{1}{10}}~,
 \end{equation}
 \end{widetext}
 is the slowly varying density correction of TM exchange energy functional~\cite{Tao-Mo16}. $\epsilon_x^{unif}=3k_f/4\pi$ is the exchange 
 energy per electron of the uniform electron gas and the terms associated with the exchange enhancement factor are $f=[1+10(70y/27)
+\beta y^2]^{1/10}$, $\mathcal{L}=[3(\lambda^2-\lambda
+1/2)(\tau-\tau^{\rm unif}-|\nabla n|^2/72n)-(\tau-\tau^{unif})+\frac{7}{18}(2\lambda-1)^2\frac{|\nabla\rho|^2}{\rho}]/
\tau^{\rm unif}$, $\mathcal{M}=(2\lambda-1)^2~p$, with $A=\frac{\mu}{2fk_f}$, $k_f=(3\pi^2\rho)^{\frac{1}{3}}$ (uniform Thomas-Fermi wave vector), 
$\tau^{\rm unif}=\frac{3}{10}k_f^2\rho$ (uniform KE density), $p=\frac{|\nabla\rho|^2}{(2k_f\rho)^2}$ (square of the reduced density 
gradient, $s=\frac{|\nabla\rho|}{(2k_f\rho)}$), $y=(2\lambda-1)^2~p$,  $\tilde{A}=\frac{\mu}{2k_f}$, $\tilde{q}=\frac{3\tau}{2(3\pi^2)^{2/3}\rho^{5/3}}
-\frac{9}{20}-\frac{p}{12}$, $\tau^w=|\nabla\rho|^2/8\rho$ and $w =[(\tau^w/\tau)^2+3(\tau^w/\tau)^3]/[1+(\tau^w/\tau)^3]^2 $ (weight factor between 
DME exchange energy and slowly varying density correction of TM functional). $E_x^{SL}$ is the full TM exchange energy functional defined 
in ref.\cite{Tao-Mo16}. For correlation part we used one electron self-interaction free TPSS ~\cite{TPSS03} correlation.
The parameter $\mu$ is fixed as the same value is used as in CAM-B3LYP functional, i.e.,~$0.33$. The mixing parameter $a$ is chosen to be $0.10$, which 
gives very good agreement to the  atomization energy of G2/148 molecular test set. It is noteworthy to mention that in the TPSSh ~\cite{VNS03}  
functional, $a=0.10$ is obtained by fitting it with molecular properties. As, our present construction is in the meta-GGA level, we stick with that 
value. Recently, proposed TPSS based RS functional by Tao et. al.~\cite{tpssrs} and HSE06~\cite{hse06} used different values of $\mu$ and $a$. Because 
those have different nature of construction compared to the present proposition. More realistic explanation of smaller mixing parameter 
in hybrid meta-GGA functional than GGA is also given in ref~\cite{VNS03}. The values of $\lambda$ and $\beta$ are the same as suggested in
TM functional~\cite{Tao-Mo16}, i.e., $0.6866$ and $79.873$.

Clearly, here we employ the full range separation scheme in the SR part of DME based exchange. For slowly varying density 
correction we employ the range separation only through the LDA exchange hole. The main motivation of employing the range separation 
only through the LDA exchange hole is rooted in the construction of the CAM-B3LYP family range separated functional~\cite{camb3lyp}, 
where the range separation is involved only through the LDA exchange hole coupled with the modified $k_f$. But in the present situation, 
we consider only $k=k_f$ in the short range semilocal part of exchange hole. Because in the solid state system, where slowly density 
varying density correction dominates, $k=k_f$ seems to be a good approximation. This is a simplified conventional way to include 
the slowly varying density scheme within the DME range separation scheme bypassing the reversed engineered exchange hole of the full 
TM exchange energy functional. Recently, the reversed engineered exchange hole for TPSS functional has been derived by Tao et. al.~\cite{tpssrs}
But the enhancement factor form of TPSS and TM functional are quite different. TPSS functional is derived from the enhancement factor of 
linear response theory. Whereas, part of TM functional is derived form spherically averaged semilocal exchange hole. This completes 
the construction of our functional.

\begin{table*}
\caption{Databases Used for benchmark calculations}
\begin{tabular}{l  c  c c c c c c c c c }
\hline\hline
 database &~~~~~~description &~~~~~~ ref. &~~~~~~ Basis set used in our present calculations\\
 \hline
 AE6                 &~~~~~~6 atomization energies&~~~~~~\cite{Quest12}&~~~~~~aug-cc-pVQZ\\
 G2                    &~~~~~~148 atomization energies&~~~~~~ ~\cite{g21,g22,g23}&~~~~~~aug-cc-pVQZ\\
 IP13                  &~~~~~~13 ionization potentials &~~~~~~ \cite{Quest12}&~~~~~~6-311++G(3df,3pd) \\
 EA13                  &~~~~~~13 electron affinities&~~~~~~ \cite{Quest12}&~~~~~~6-311++G(3df,3pd)\\ 
 PA8                   &~~~~~~8 proton affinities &~~~~~~ \cite{Quest12}&~~~~~~6-311++G(3df,3pd)\\
 HTBH38                &~~~~~~38 hydrogen transfer barrier height&~~~~~~ \cite{Quest12}&~~~~~~aug-cc-pVQZ\\
 NHTBH38               &~~~~~~38 non-hydrogen transfer barrier height&~~~~~~ \cite{Quest12}&~~~~~~aug-cc-pVQZ\\
 $\pi$TC13             &~~~~~~13 thermochemistry of $\pi$ system&~~~~~~ \cite{Quest12}&~~~~~~6-311+G(3df,3pd)\\
 ABDE12                &~~~~~~12 alkyl bond dissociation energies&~~~~~~ \cite{Quest12}&~~~~~~aug-cc-pVQZ \\
 HC7                   &~~~~~~7 Hydrocarbon chemistry &~~~~~~ \cite{Quest12}&~~~~~~6-311++G(3df,3pd) \\ 
 ISOL6                 &~~~~~~6 isomerization energies &~~~~~~ \cite{Quest12}&~~~~~~6-311++G(3df,3pd)\\
 \hline\hline
\label{tab1}
\end{tabular}
\end{table*}


\begin{table*}
\caption{Summary of deviation using different methods. Mean error (ME) and Mean absolute error (MAE) are calculated here. 
The basis set used in all the benchmark calculations are summarized in Table-(\ref{tab1})}
\centering
  \begin{adjustbox}{max width=\textwidth}
  \begin{tabular}{*{12}{|c}|}
\hline \hline  
  &\multirow{2}{40pt}{AE6~~G2}&
  \multirow{2}{40pt}{~~ABDE} &
  \multirow{2}{20pt}{~~HC} &
  \multirow{2}{20pt}{~EA} &
  \multirow{2}{20pt}{~~IP} &
  \multirow{2}{40pt}{~~~~PA}&
  \multirow{2}{20pt}{IsoL} &
  \multirow{2}{20pt}{$\pi$TC} &
  \multirow{2}{20pt}{BH} \\
   &    &  & & & & & & &  \\ 
    &Kcal/mol    & Kcal/mol& Kcal/mol & eV & eV & eV & Kcal/mol& Kcal/mol& Kcal/mol\\ 
    &    &  & & & & & & &  \\
Functionals    &MAE~   MAE &ME~  MAE &ME~  MAE &ME~  MAE&
 ME~  MAE&ME~ MAE &ME~  MAE &ME~  MAE &ME~  MAE \\ \hline

B3LYP &2.7~~~~3.599   &9.91~~9.91&15.93~~15.93& -0.061~~0.095&-0.075~~0.227   &-0.012~~0.047  &2.54~~2.54&-5.64~~5.77&2.81~~5.08\\
PBE0 &5.8~~~~5.619 &7.26~~7.26&-5.34~~10.07& 0.065~~0.120     &-0.107~~0.137  &-0.052~~0.053     &0.58~~1.44&-5.80~~5.89&2.12~~4.70  \\
TPSSh  &6.3~~~~5.296&10.73~~10.73&5.69~~6.31&0.065~~0.122 &-0.085~~0.136  &-0.122~~0.122   &3.10~~3.10    &-7.64~~7.82 & 4.88~~6.78\\
HSE06 &6.7~~~~5.335&9.73~~9.73&0.14~~5.92&-0.068~~0.123   &0.108~~0.139  &0.077~~0.077 &-1.12~~1.42 &6.44~~6.54 &1.46~~4.08 \\
PW-TPSSc&5.7~~~~4.008&7.93 7.93&2.24~~3.21& 0.145~0.138 &-0.010~~0.098  &-0.137~0.137&2.42~~2.42&-7.93~~8.17&3.55~5.54 \\ 
\hline\hline
\end{tabular}
\end{adjustbox}
\label{tab2}
\end{table*}

The self-consistence benchmark calculation of the newly constructed range separated functional is performed using the NWChem-6.6 \cite{nwchem} code. 
The data set we used in our calculation is summarized in Table-(\ref{tab1}). We compared our results with that of three popular hybrid functional B3LYP, 
PBE0 and TPSSh and one RS hybrid functional HSE06. We choose only one RS functional i.e., HSE06 in our comparison because both formalism are constructed 
using short range HF. In Table-(\ref{tab2}), we reported the mean absolute error (ME) and mean absolute error (MAE) for all the test cases except atomization 
energy, where only MAE is reported. The MAE and ME of our proposed functional are given at the last row of Table - (\ref{tab1}). We called it 
PW-TPSSc (where PW stands for present work and it is coupled with TPSS correlation).

{\it{\textbf{Atomization energy:}}}~~The atomization energy is defined as the energy required to isolate the constituent atoms from its molecular structure. 
For atomization energy benchmark we used G2 set, which consists of 148 molecules and the geometries are optimized in MP2/6-31G* level~\cite{g21,g22,g23}. 
We used aug-cc-pVQZ basis set in our calculation. The reference values are taken from CCSD(T) calculation given in ref.~\cite{ccsdt}. Table (\ref{tab2}) 
implies that, B3LYP gives the smallest MAE both for AE6~\cite{Quest12} and G2 set and it is not surprising because B3LYP is designed to minimize the 
atomization error. PW-TPSSc is the second bast with an MAE $4.008$ and $5.7$ for G2 and AE6. For AE6 test set, PW-TPSSc perform better than 
HSE06. Among all the functionals we tested here, PBE0 has the most MAE for G2.

{\it{\textbf{Ionization potential (IP), electron affinity (EA) and Proton affinity (PA)}}:}~~IP and EA are calculated using Minnesota 2.0 data set~\cite{Quest12} with 
QCISD/MG3 level optimized geometries. For IP, PW-TPSSc performs best with smallest MAE of 0.098 eV. For the EA13 test set performance of B3LYP is superior to other 
functionals under consideration. Other functionals perform almost equivalently in this case. Proton affinity (PA) is the amount of energy released when a 
proton is added to a species at its ground state. PAs~\cite{pa1,pa2} are calculated using MP2/6-31G(2df,p) level optimized geometry. Among all the functionals, 
reported here, B3LYP is the best with MAE 0.047 eV. In this case, the performance of HSE06 is better than the present proposition. 

{\it{\textbf{Alkyl bond dissociation energies, Hydrocarbon chemistry, Isomerization energies of large molecules and Thermochemistry of $\pi$ systems}}:}~~Alkyl 
bond dissociation energy database (ABDE12)~\cite{Quest12} contains 12 molecules that include four bond dissociation energies of methyl, isopropyl, CH$_3$ and 
OCH$_3$ and another eight molecules formed from ethyl, tert-butyl, H, CH$_3$, OCH$_3$ and OH. The performance of PBE0 is best with MAE 7.268 Kcal/mol 
compared to others. Next, best performance is observed for PW-TPSSc with MAE $7.932$ Kcal/mol. In case of hydrocarbon chemistry, ~\cite{Quest12} PW-TPSSc 
outperformed all other functionals at least MAE $3.217$ Kcal/mol. For IsoL6 data set, HSE06 achieves smallest MAE, while the performance of PW-TPSSc is better 
than TPSSh meta-GGA functional. For $\pi$ system we have considered MP2/6-31+G(d,p) level optimized geometries taken from the Minnesota 2.0 database~\cite{Quest12}.  
The B3LYP gives the smallest MAE with 5.776 Kcal/mol. Whereas, largest MAE is obtained from PW-TPSSc with MAE 8.177 Kcal/mol.

{\it{\textbf{Barrier heights of chemical reactions}}:}~~ Due to the transition states with stretched bonds the reaction barrier heights of chemical reactions are 
related to the many electron self interaction error. Therefore, functional with least MESI always perform well for barrier heights. Semilocal functionals with long 
range correction always perform well in predicting barrier heights. Geometries and the corresponding reference values are again taken from the Minnesota 
2.0 database~\cite{Quest12}. The data set consists of forward and reverse barrier heights of 19 hydrogen and 19 non-hydrogen transfer reactions. 
aug-cc-pVQZ basis set has been used for all our functionals calculations. From Table - (\ref{tab2}), HSE06 performs gives the lowest MAE for the 
76 test set with MAE 4.084 Kcal/mol. Performance of PW-TPSSc is better than TPSSh. The reason behind the performance of HSE06 is better than 
PW-TPSSc is due to the fact that HSE06 have $1/4$ fraction exact exchange with its semilocal form, therefore, more MESI free. B3LYP and PBE0 
perform equivalently for the BH76 test set with MAE 5.083 Kcal/mol and 4.709 Kcal/mol.

\begin{figure}
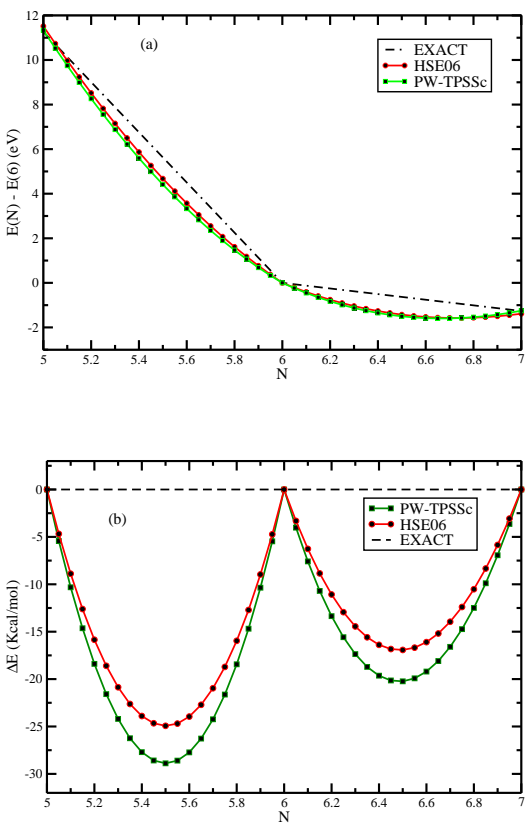

\includegraphics[width=.80\linewidth]{comp-tmx-hse06.eps}\\
\vspace{1cm}
\includegraphics[width=.80\linewidth]{Carbon-b-plot.eps}
\caption{ (a) Shown is the energy difference of C atom with respect to fractional electron occupation number for HSE06 and present case. The exact 
line is obtained using experimental -IP and -EA.(b) The deviation of HSE06 and present proposition from the piece-wise linear extrapolation with 
respect to fractional electron occupation number. The 6-311++(3df,3pd) basis set is used for all the calculations.}
\label{fig1}
\end{figure}

Lastly, we want to draw a comparison of the behavior of HSE06 and PW-TPSSc in case of fractional electron occupation number of C atom. this phenomenon 
is directly related to the semiconductor band gap~\cite{pplb82}. We give special interest to these two functionals because both are designed considering 
HF in its short range part. Comparison from Fig-(\ref{fig1}) clearly indicated that both perform equivalently. HSE06 have $1/4$ HF mixing in its SR part 
compared to the PW-TPSSc, that's why it behaves slightly better. This difference is only evident when we plot them in Kcal/mol scale as it is shown in Fig-(\ref{fig1}) (b).      

A meta-GGA level range separated hybrid functional is proposed using exact HF exchange in its short range. The RS functional proposed here utilizes the 
full DME based exchange hole coupled with the slowly varying density correction of the enhancement factor included through the LDA exchange hole. This 
is the first ever attempt to utilize the TM functional in RS perspective. Comprehensive assessment of present functional with B3LYP, PBE0, TPSSh and 
HSE06 shows it performs promisingly in several cases. Specially. G2/148 , ABDE12, HC7 and IP13 test cases performance of PW-TPSSc is quite impressive.  
Performance of HSE06 and PW-TPSSc for the fractional occupation number perspective has also been discussed because both are designed using HF in short range. 
It has been observed that both perform almost similarly in fractional change prospective. Lastly, we conclude that, the functional we developed here can be 
used further in solid state calculations. Further extensions and performance to the solid state system of the proposed functional will be reported in future.           


\end{document}